\documentclass{article}
\usepackage{spconf,amsmath,graphicx}
\usepackage{textcomp}
\usepackage{xcolor} 

\title{Speech Synthesis using EEG}
\name{Gautam Krishna \qquad Co Tran \qquad Yan Han$^{\star}$ \qquad Mason Carnahan$^{\star}$ \qquad Ahmed H Tewfik \thanks{$\star$ Equal author contribution}}
\address{Brain Machine Interface Lab, The University
of Texas at Austin \\}
\begin{document}
%
\maketitle
\begin{abstract}
In this paper we demonstrate speech synthesis using different electroencephalography (EEG) feature sets recently introduced in \cite{krishna2019state}.
We make use of a recurrent neural network (RNN) regression model to predict acoustic features directly from EEG features. We demonstrate our results using EEG features recorded in parallel with spoken speech as well as using EEG recorded in parallel with listening utterances. We provide EEG based speech synthesis results for four subjects in this paper and our results demonstrate the feasibility of synthesizing speech directly from EEG features. 
\end{abstract}
\begin{keywords}
Speech synthesis, EEG, Deep Learning
\end{keywords}
\section{Introduction}
\label{sec:intro}

Speech production is one of the most important abilities of human beings which helps humans to communicate with each other. 
In recent years there is lot of research interest in developing assistive technologies to help with speech restoration for people with speaking disabilities. In \cite{anumanchipalli2019speech} authors demonstrated synthesizing speech directly from invasive electrocorticography (ECoG) neural recordings using a recurrent neural network (RNN) based speech decoder. Recently in \cite{krishna2019state,krishna2019speech,krishna20} authors have demonstrated speech recognition using electroencephalography (EEG) features where EEG signals recorded in parallel with spoken speech \cite{krishna20,krishna2019speech} as well as EEG signals recorded in parallel with listening utterances \cite{krishna2019state} are translated directly into text. EEG is a non invasive way of measuring electrical activity of human brain. EEG has high temporal resolution like the invasive ECoG signals. The subjects need not undergo a brain surgery like in the case of ECoG for recording EEG since EEG is completely a non invasive technique. EEG sensors are placed on the scalp of the subjects to obtain the recordings. 

In \cite{krishna2019state} authors introduced three types of EEG feature sets which are useful for speech recognition and speech synthesis. Though in \cite{krishna2019state} authors demonstrated preliminary results for synthesizing speech directly from EEG signals using a long short term memory (LSTM) \cite{hochreiter1997long} based regression model and generative adversarial network (GAN) \cite{goodfellow2014generative} based model, they didn't provide speech synthesis results per each subject for their experiments and they provided only EEG feature set 1 based result for synthesizing speech from EEG signals recorded in parallel with spoken speech. Where as in this paper we provide speech synthesis results per each subject using all the three different EEG feature sets for EEG signals recorded in parallel with spoken speech as well as from EEG signals recorded in parallel with listening utterances. More over in this paper we introduce a new EEG data set consisting of EEG, speech signals recorded for commonly used voice commands and our overall results mentioned in this paper demonstrate a significant improvement compared to the results demonstrated for speech synthesis by authors in \cite{krishna2019state}. We demonstrate in this paper that using a gated recurrent unit (GRU) \cite{chung2014empirical} based RNN network with dropout \cite{srivastava2014dropout} instead of a LSTM network results in speech synthesis performance improvement as well as in this paper we perform speech synthesis experiments using more number of speech, EEG recording examples per sentence compared to the ones used by authors in \cite{krishna2019state}. 

For synthesizing speech from EEG signals recorded in parallel with spoken speech, with our approach we were able to achieve a mel cepstral distortion (MCD) \cite{kominek2008synthesizer} value as low as 0.433 compared to 5.737 demonstrated by authors in \cite{krishna2019state} and for synthesizing speech from EEG signals recorded in parallel with listening utterances, with our approach we were able to achieve a MCD value as low as 0.471 compared to 1.34 demonstrated by authors in \cite{krishna2019state}. Given the challenges outlined by authors in \cite{krishna2019state} for training a GAN model for speech synthesis using EEG, we didn't perform experiments with GAN model. 

Results mentioned in this paper demonstrates a significant first step towards synthesizing speech directly from EEG features. Synthesizing speech directly from EEG signals might help amyotrophic lateral sclerosis (ALS) patients who lost ability to speak with speech restoration.

\section{Speech Synthesis Model}
\label{sec:format}

Our speech synthesis model consists of two layers of gated recurrent unit (GRU) \cite{chung2014empirical} with 256 hidden units in first layer and 128 hidden units in second layer as shown in Figure 1. The final GRU layer is connected to a time distributed dense layer of 13 hidden units to predict acoustic features of dimension 13 at every time step. Between each GRU layers and between the final GRU layer and time distributed dense layer, a dropout regularization \cite{srivastava2014dropout} with dropout rate 0.2 is applied. The model takes EEG features as input at every time step and outputs acoustic features at every time step. The model was trained for 250 epochs with adam optimizer with learning rate 0.01 \cite{kingma2014adam} to observe loss convergence. The batch size was set to 100 and mean squared error (MSE) was used as the regression loss function for the model. For each subject's data we used 80 \% data to train the model and remaining 10 \% for testing and rest 10 \% for validation set.
The validation set was used to identify the right values for hyper parameters for the model. All the scripts for the model were written using Keras deep learning python framework. 

\begin{figure}[h]
\label{fig:asrmodel}
\includegraphics[height=8.5cm, width=\linewidth,trim={0.1cm 0.1cm 0.1cm 0.1cm}]{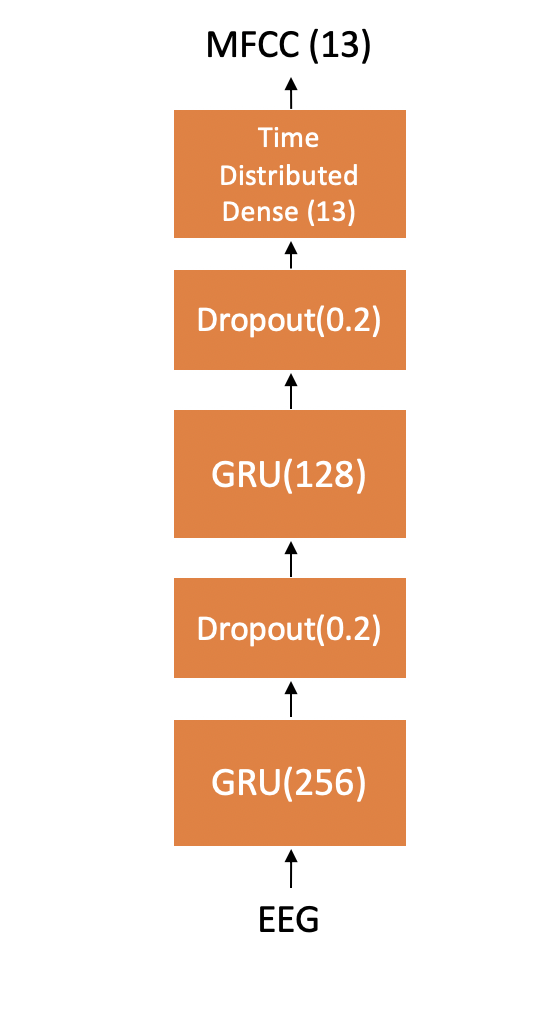}
\caption{Speech Synthesis Model} 
\label{1vsall}
\end{figure}

\section{Design of Experiments for building the database}
\label{sec:pagestyle}

Four subjects took part in the EEG recording experiments. All were UT Austin undergraduate students in their early twenties. Three were females and one subject was male. Each subject was first asked to listen to four different natural utterances and then speak out loud the utterances that they listened to. The EEG was recorded in parallel while they were listening to the utterances as well as EEG was recorded in parallel while they were speaking out the utterances that they listened to. Throughout this paper we will refer to the EEG recorded in parallel with listening as listen EEG and EEG recorded in parallel with speech as spoken EEG. The listening utterances were simultaneously recorded with listen EEG and the subject's speech was recorded simultaneously with spoken EEG.
The four natural utterances that the subjects listened were "Hi Bixby", "Call Mom", "Open Camera" and " What's the weather".
We collected 70 Speech - EEG recordings per each subject per each sentence. 

We used Brain Vision EEG recording hardware. Our EEG cap had 32 wet EEG electrodes including one electrode as \textbf{ground} as shown in Figure 2. We used EEGLab \cite{delorme2004eeglab} to obtain the EEG sensor location mapping. It is based on standard 10-20 EEG sensor placement method for 32 electrodes \cite{sharbrough1991american}.

\begin{figure}[h]
\begin{center}
\includegraphics[height=3cm,width=0.25\textwidth,trim={1cm 1cm 1cm 0.1cm},clip]{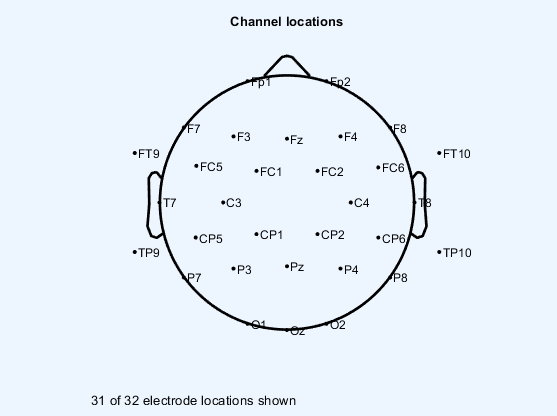}
\caption{EEG channel locations for the cap used in our experiments} 
\label{1vsall}
\end{center}
\end{figure}

\section{EEG and Speech feature extraction details}
\label{sec:typestyle}

EEG signals were sampled at 1000Hz and a fourth order IIR band pass filter with cut off frequencies 0.1Hz and 70Hz was applied. A notch filter with cut off frequency 60 Hz was used to remove the power line noise.
EEGlab's \cite{delorme2004eeglab} Independent component analysis (ICA) toolbox was used to remove other biological signal artifacts like electrocardiography (ECG), electromyography (EMG), electrooculography (EOG) etc from the EEG signals. 
We then extracted the three EEG feature sets explained by authors in \cite{krishna2019state}. The details of each EEG feature set are covered in \cite{krishna2019state}. Each EEG feature set was extracted at a sampling frequency of 100 Hz for each EEG channel \cite{krishna2019speech}. 

The recorded speech signal was sampled at 16KHz frequency. We extracted mel-frequency cepstral coefficients (MFCC) as features for speech signal. We extracted MFCC features of dimension 13.
The MFCC features were also sampled at 100Hz same as the sampling frequency of EEG features. 

\section{EEG Feature Dimension Reduction Algorithm Details}
\label{sec:majhead}

We used kernel principal component analysis (KPCA) \cite{mika1999kernel} to de-noise the EEG feature space by performing dimension reduction for each EEG feature set.
By following the dimension reduction methods explained by authors in \cite{krishna2019state} we reduced EEG feature set 1 to a dimension of 30, EEG feature set 2 was reduced to a dimension of 50 and EEG feature set 3 was kept at original dimension of 93. More details of explained variance plots used to identify the right feature dimensions are covered in \cite{krishna2019state}.

\section{Results}
\label{sec:print}

Like mentioned earlier, for each subject we used 10 \% of data as test set. During test time, EEG features from test set are fed as input to the trained speech synthesis model to output MFCC or acoustic features. 

We computed three types of performance metrics namely the mel cepstral distortion (MCD) \cite{kominek2008synthesizer}, root mean squared error (RMSE) and normalized RMSE between the predicted MFCC during test time and ground truth MFCC from test set to evaluate the performance of the model on test set for each subject. The RMSE values were normalized by dividing the RMSE values with the absolute difference between the maximum and minimum value in the test set observation vector. The predicted and ground truth MFCC values were normalized before computing the MCD values.

Tables 1,2,3 and 4 shows results obtained during test time for predicting listen MFCC features from listen EEG features for each of the four subjects for various EEG feature set inputs. 
Tables 5,6,7 and 8 shows results obtained during test time for predicting spoken MFCC features from spoken EEG features for each of the four subjects for various EEG feature set inputs. 

We used Griffin Lim reconstruction \cite{griffin1984signal} algorithm to convert the predicted test time MFCC or acoustic features to audio or speech waveforms. We observed comparable MCD values for each subject for different EEG feature sets for each of the speech synthesis experiment as seen from the results shown in tables. Our results indicate if sufficient amount of data is available for training the speech synthesis model, then the effect on choice of EEG feature sets is negligible. 

However we observed some what significant variation in MCD values across different subjects which indicates each brain generates unique set of signals during speech perception and production even though the different subjects were listening to the same natural utterances and speaking out loud the same sentences.

\begin{table}[!ht]
\centering
\begin{tabular}{|l|l|l|l|}
\hline
\textbf{\begin{tabular}[c]{@{}l@{}}EEG\\ Feature\\ Set\end{tabular}} & \textbf{\begin{tabular}[c]{@{}l@{}}Average\\ MCD\end{tabular}} & \textbf{\begin{tabular}[c]{@{}l@{}}Average\\ RMSE\end{tabular}} & \textbf{\begin{tabular}[c]{@{}l@{}}Average\\ Normalized\\ RMSE\end{tabular}} \\ \hline
Set 1                                                                & 0.4713                                                         & 5.557                                                           & 0.0114                                                                       \\ \hline
Set 2                                                                & 0.4723                                                         & 5.540                                                           & 0.0114                                                                       \\ \hline
Set 3                                                                & 0.4744                                                         & 5.536                                                           & 0.0113                                                                       \\ \hline
\end{tabular}
\caption{Results for predicting \textbf{listen MFCC from listen EEG for subject 1}}
\end{table}

\begin{table}[!ht]
\centering
\begin{tabular}{|l|l|l|l|}
\hline
\textbf{\begin{tabular}[c]{@{}l@{}}EEG\\ Feature\\ Set\end{tabular}} & \textbf{\begin{tabular}[c]{@{}l@{}}Average\\ MCD\end{tabular}} & \textbf{\begin{tabular}[c]{@{}l@{}}Average\\ RMSE\end{tabular}} & \textbf{\begin{tabular}[c]{@{}l@{}}Average\\ Normalized\\ RMSE\end{tabular}} \\ \hline
Set 1                                                                & 0.829                                                          & 7.636                                                           & 0.0225                                                                        \\ \hline
Set 2                                                                & 0.80                                                           & 7.419                                                           & 0.0227                                                                        \\ \hline
Set 3                                                                & 0.814                                                          & 7.521                                                           & 0.0221                                                                        \\ \hline
\end{tabular}
\caption{Results for predicting \textbf{listen MFCC from listen EEG for subject 2}}
\end{table}

\begin{table}[!ht]
\centering
\begin{tabular}{|l|l|l|l|}
\hline
\textbf{\begin{tabular}[c]{@{}l@{}}EEG\\ Feature\\ Set\end{tabular}} & \textbf{\begin{tabular}[c]{@{}l@{}}Average\\ MCD\end{tabular}} & \textbf{\begin{tabular}[c]{@{}l@{}}Average\\ RMSE\end{tabular}} & \textbf{\begin{tabular}[c]{@{}l@{}}Average\\ Normalized\\ RMSE\end{tabular}} \\ \hline
Set 1                                                                & 0.64                                                           & 6.69                                                            & 0.016                                                                        \\ \hline
Set 2                                                                & 0.638                                                          & 6.70                                                            & 0.016                                                                        \\ \hline
Set 3                                                                & 0.63                                                           & 6.69                                                            & 0.016                                                                        \\ \hline
\end{tabular}
\caption{Results for predicting \textbf{listen MFCC from listen EEG for subject 3}}
\end{table}

\begin{table}[!ht]
\centering
\begin{tabular}{|l|l|l|l|}
\hline
\textbf{\begin{tabular}[c]{@{}l@{}}EEG\\ Feature\\ Set\end{tabular}} & \textbf{\begin{tabular}[c]{@{}l@{}}Average\\ MCD\end{tabular}} & \textbf{\begin{tabular}[c]{@{}l@{}}Average\\ RMSE\end{tabular}} & \textbf{\begin{tabular}[c]{@{}l@{}}Average\\ Normalized\\ RMSE\end{tabular}} \\ \hline
Set 1                                                                & 1.759                                                          & 13.11                                                           & 0.05                                                                       \\ \hline
Set 2                                                                & 1.765                                                          & 13.14                                                           & 0.05                                                                        \\ \hline
Set 3                                                                & 1.758                                                          & 13.12                                                           & 0.05                                                                        \\ \hline
\end{tabular}
\caption{Results for predicting \textbf{listen MFCC from listen EEG for subject 4}}
\end{table}

\begin{table}[!ht]
\centering
\begin{tabular}{|l|l|l|l|}
\hline
\textbf{\begin{tabular}[c]{@{}l@{}}EEG\\ Feature\\ Set\end{tabular}} & \textbf{\begin{tabular}[c]{@{}l@{}}Average\\ MCD\end{tabular}} & \textbf{\begin{tabular}[c]{@{}l@{}}Average\\ RMSE\end{tabular}} & \textbf{\begin{tabular}[c]{@{}l@{}}Average\\ Normalized\\ RMSE\end{tabular}} \\ \hline
Set 1                                                                & 0.433                                                          & 4.867                                                           & 0.0105                                                                       \\ \hline
Set 2                                                                & 0.435                                                          & 4.881                                                           & 0.0105                                                                       \\ \hline
Set 3                                                                & 0.435                                                          & 4.911                                                           & 0.0106                                                                       \\ \hline
\end{tabular}
\caption{Results for predicting \textbf{spoken MFCC from spoken EEG for subject 1}}
\end{table}

\begin{table}[!ht]
\centering
\begin{tabular}{|l|l|l|l|}
\hline
\textbf{\begin{tabular}[c]{@{}l@{}}EEG\\ Feature\\ Set\end{tabular}} & \textbf{\begin{tabular}[c]{@{}l@{}}Average\\ MCD\end{tabular}} & \textbf{\begin{tabular}[c]{@{}l@{}}Average\\ RMSE\end{tabular}} & \textbf{\begin{tabular}[c]{@{}l@{}}Average\\ Normalized\\ RMSE\end{tabular}} \\ \hline
Set 1                                                                & 0.856                                                          & 8.04                                                            & 0.0237                                                                        \\ \hline
Set 2                                                                & 0.847                                                          & 8.03                                                            & 0.0236                                                                        \\ \hline
Set 3                                                                & 0.841                                                          & 7.96                                                            & 0.0232                                                                        \\ \hline
\end{tabular}
\caption{Results for predicting \textbf{spoken MFCC from spoken EEG for subject 2}}
\end{table}

\begin{table}[!ht]
\centering
\begin{tabular}{|l|l|l|l|}
\hline
\textbf{\begin{tabular}[c]{@{}l@{}}EEG\\ Feature\\ Set\end{tabular}} & \textbf{\begin{tabular}[c]{@{}l@{}}Average\\ MCD\end{tabular}} & \textbf{\begin{tabular}[c]{@{}l@{}}Average\\ RMSE\end{tabular}} & \textbf{\begin{tabular}[c]{@{}l@{}}Average\\ Normalized\\ RMSE\end{tabular}} \\ \hline
Set 1                                                                & 0.647                                                          & 6.442                                                           & 0.0155                                                                        \\ \hline
Set 2                                                                & 0.650                                                          & 6.43                                                            & 0.0156                                                                         \\ \hline
Set 3                                                                & 0.645                                                          & 6.437                                                           & 0.0156                                                                        \\ \hline
\end{tabular}
\caption{Results for predicting \textbf{spoken MFCC from spoken EEG for subject 3}}
\end{table}

\begin{table}[!ht]
\centering
\begin{tabular}{|l|l|l|l|}
\hline
\textbf{\begin{tabular}[c]{@{}l@{}}EEG\\ Feature\\ Set\end{tabular}} & \textbf{\begin{tabular}[c]{@{}l@{}}Average\\ MCD\end{tabular}} & \textbf{\begin{tabular}[c]{@{}l@{}}Average\\ RMSE\end{tabular}} & \textbf{\begin{tabular}[c]{@{}l@{}}Average\\ Normalized\\ RMSE\end{tabular}} \\ \hline
Set 1                                                                & 1.733                                                          & 13.19                                                           & 0.053                                                                        \\ \hline
Set 2                                                                & 1.736                                                          & 13.19                                                           & 0.054                                                                        \\ \hline
Set 3                                                                & 1.741                                                          & 13.23                                                           & 0.053                                                                        \\ \hline
\end{tabular}
\caption{Results for predicting \textbf{spoken MFCC from spoken EEG for subject 4}}
\end{table}


\section{conclusion}
\label{sec:refs}
In this paper we demonstrated synthesizing speech directly from non invasive EEG neural recording using a simple GRU based regression model for four subjects using different types of EEG feature sets. For synthesizing speech from EEG signals recorded in parallel with spoken speech, with our approach we were able to achieve a mel cepstral distortion (MCD) value as low as 0.433 compared to 5.737 demonstrated by authors in \cite{krishna2019state} and for synthesizing speech from EEG signals recorded in parallel with listening utterances, with our approach we were able to achieve a MCD value as low as 0.471 compared to 1.34 demonstrated by authors in \cite{krishna2019state}. Our results demonstrate adding regularization factors like dropouts to the speech synthesis model and training with more number of examples per sentence helps in reducing the test time MCD values. 
Our results demonstrates a significant first step towards synthesizing speech directly from EEG features. We further plan to publish the data sets used in this work to help advancement of research.

\section{Acknowledgement}
We would like to thank Kerry Loader and Rezwanul Kabir from Dell, Austin, TX for donating us the GPU to train the models used in this work.
\bibliographystyle{IEEEbib}
\bibliography{strings,refs}

\begin{thebibliography}{10}

\bibitem{krishna2019state}
Gautam Krishna, Yan Han, Co~Tran, Mason Carnahan, and Ahmed~H Tewfik,
\newblock ``State-of-the-art speech recognition using eeg and towards decoding
  of speech spectrum from eeg,''
\newblock {\em arXiv preprint arXiv:1908.05743}, 2019.

\bibitem{anumanchipalli2019speech}
Gopala~K Anumanchipalli, Josh Chartier, and Edward~F Chang,
\newblock ``Speech synthesis from neural decoding of spoken sentences,''
\newblock {\em Nature}, vol. 568, no. 7753, pp. 493, 2019.

\bibitem{krishna2019speech}
Gautam Krishna, Co~Tran, Jianguo Yu, and Ahmed Tewfik,
\newblock ``Speech recognition with no speech or with noisy speech,''
\newblock in {\em Acoustics, Speech and Signal Processing (ICASSP), 2019 IEEE
  International Conference on}. IEEE, 2019.

\bibitem{krishna20}
Gautam Krishna, Co~Tran, Mason Carnahan, and Ahmed Tewfik,
\newblock ``Advancing speech recognition with no speech or with noisy speech,''
\newblock in {\em 2019 27th European Signal Processing Conference (EUSIPCO)}.
  IEEE, 2019.

\bibitem{hochreiter1997long}
Sepp Hochreiter and J{\"u}rgen Schmidhuber,
\newblock ``Long short-term memory,''
\newblock {\em Neural computation}, vol. 9, no. 8, pp. 1735--1780, 1997.

\bibitem{goodfellow2014generative}
Ian Goodfellow, Jean Pouget-Abadie, Mehdi Mirza, Bing Xu, David Warde-Farley,
  Sherjil Ozair, Aaron Courville, and Yoshua Bengio,
\newblock ``Generative adversarial nets,''
\newblock in {\em Advances in neural information processing systems}, 2014, pp.
  2672--2680.

\bibitem{chung2014empirical}
Junyoung Chung, Caglar Gulcehre, KyungHyun Cho, and Yoshua Bengio,
\newblock ``Empirical evaluation of gated recurrent neural networks on sequence
  modeling,''
\newblock {\em arXiv preprint arXiv:1412.3555}, 2014.

\bibitem{srivastava2014dropout}
Nitish Srivastava, Geoffrey Hinton, Alex Krizhevsky, Ilya Sutskever, and Ruslan
  Salakhutdinov,
\newblock ``Dropout: a simple way to prevent neural networks from
  overfitting,''
\newblock {\em The journal of machine learning research}, vol. 15, no. 1, pp.
  1929--1958, 2014.

\bibitem{kominek2008synthesizer}
John Kominek, Tanja Schultz, and Alan~W Black,
\newblock ``Synthesizer voice quality of new languages calibrated with mean mel
  cepstral distortion,''
\newblock in {\em Spoken Languages Technologies for Under-Resourced Languages},
  2008.

\bibitem{kingma2014adam}
Diederik~P Kingma and Jimmy Ba,
\newblock ``Adam: A method for stochastic optimization,''
\newblock {\em arXiv preprint arXiv:1412.6980}, 2014.

\bibitem{delorme2004eeglab}
Arnaud Delorme and Scott Makeig,
\newblock ``Eeglab: an open source toolbox for analysis of single-trial eeg
  dynamics including independent component analysis,''
\newblock {\em Journal of neuroscience methods}, vol. 134, no. 1, pp. 9--21,
  2004.

\bibitem{sharbrough1991american}
Frank Sharbrough,
\newblock ``American electroencephalographic society guidelines for standard
  electrode position nomenclature,''
\newblock {\em J clin Neurophysiol}, vol. 8, pp. 200--202, 1991.

\bibitem{mika1999kernel}
Sebastian Mika, Bernhard Sch{\"o}lkopf, Alex~J Smola, Klaus-Robert M{\"u}ller,
  Matthias Scholz, and Gunnar R{\"a}tsch,
\newblock ``Kernel pca and de-noising in feature spaces,''
\newblock in {\em Advances in neural information processing systems}, 1999, pp.
  536--542.

\bibitem{griffin1984signal}
Daniel Griffin and Jae Lim,
\newblock ``Signal estimation from modified short-time fourier transform,''
\newblock {\em IEEE Transactions on Acoustics, Speech, and Signal Processing},
  vol. 32, no. 2, pp. 236--243, 1984.

\end{thebibliography}

\end{document}